\documentclass[twocolumn]{aa}
\usepackage{amsmath}
\usepackage{amssymb}
\usepackage{graphicx}
\usepackage{txfonts}
\usepackage{hyperref}  
\usepackage{multirow}
\usepackage{float}
\usepackage{orcidlink}
\usepackage{soul}
\usepackage{xcolor}
\hypersetup{colorlinks = true, allcolors = blue}
\usepackage{CJKutf8}

\usepackage{placeins}           

\begin{document} 
   \title{Search for quasar pairs with \textit{Gaia} astrometric data}
   \subtitle{IV. Confirmation of 17 dual quasars and 143 projected quasars}
   \titlerunning{Search for quasar pairs with \textit{Gaia} astrometric data. IV.}
   \author{
      Zhuojun Deng (\begin{CJK}{UTF8}{gbsn}{邓卓君}\end{CJK}) 
      \inst{1,2}
      \orcidlink{0009-0008-8080-3124}
      \and
      Qihang Chen (\begin{CJK}{UTF8}{gbsn}{陈启航}\end{CJK}) \inst{1,2}\orcidlink{0009-0006-9345-9639}
      \and
      Liang Jing (\begin{CJK}{UTF8}{gbsn}{荆亮}\end{CJK})
      \inst{1,2}\orcidlink{0000-0003-1188-9573}
      \and \\
      Xingyu Zhu (\begin{CJK}{UTF8}{gbsn}{朱星宇}\end{CJK})
      \inst{1,2}\orcidlink{0009-0008-9072-4024}
      \and
      Jianghua Wu (\begin{CJK}{UTF8}{gbsn}{吴江华}\end{CJK}) \inst{1,2} \thanks{Corresponding authors: \href{mailto:jhwu@bnu.edu.cn}{jhwu@bnu.edu.cn}} \orcidlink{0000-0002-8709-6759}}
    
    \institute{
                School of Physics and Astronomy, Beijing Normal University, Beijing, 100875, China\\ 
      \email{jhwu@bnu.edu.cn}
      \and
      Institute for Frontier in Astronomy and Astrophysics, Beijing Normal University, Beijing, 102206, China}
      

  \abstract 
    {Dual quasars separated at kiloparsec scale are widely regarded as precursors to binary supermassive black holes and offer a key insight into the dynamical evolution of galaxy mergers. Our series of studies focuses on searching for dual quasars by using a selection strategy of zero proper motion and zero parallax to isolate candidate quasars near known ones and by follow-up spectroscopy of the candidates. This paper, the fourth in the series, reports the spectroscopic confirmations of our quasar pair candidates based on the spectroscopic data of the SDSS and DESI DR1. We newly identified 17 dual quasars and 143 projected quasars.  The redshifts of the 17 dual quasars range from 0.573 to 2.758, with a median of 1.512. One notable system, J0023+0417, exhibits nearly identical spectral features in the two members and shows evidence of a potential foreground galaxy, making it a high-confidence strong gravitational lensing system. The redshifts of the 143 projected quasars are from 0.301 to 4.030, with a median of 1.596. Among them, four have projected distances below 30 kpc, offering valuable opportunities to probe the circumgalactic medium (CGM) of the foreground host galaxy through absorption lines.}
   

   \maketitle
%

    \section{Introduction} \label{sec:intro}
    Dark matter halos form through hierarchical structure formation within the prevailing $\Lambda$CDM cosmological model. Small-scale dark matter halos first undergo gravitational collapse, and subsequently grow into massive systems through hierarchical mergers, thereby driving galaxy evolution \citep{1978MNRAS.183..341W, Begelman1980Nature}. Within this framework, galaxy mergers emerge as a generic consequence of the ongoing hierarchical mass assembly \citep{Sanders1996, 2000MNRAS.319..168C, Hopkins2008, Sandrinelli2018QPenvironment}. 

    Supermassive black holes (SMBHs; $>10^8\,M_\odot$) are believed to be ubiquitous in the centres of massive galaxies \citep{Richstone1998, Kormendy2013}. 
    A number of theoretical and observational works suggest that galaxy interactions and mergers can drive gas inflows and trigger accretion activity \citep{Bellovary2013ApJ, Goulding2018, Omori2025ApJ}. When both SMBHs are active, these systems can be observed as dual AGNs or dual quasars, with transverse separations ranging from 30 pc to 110 kpc \citep[e.g.,][]{Pfeifle2025ApJS}.

    As the merger proceeds, the two SMBHs sink towards the centre through dynamical friction \citep{Begelman1980Nature, Kormendy2013} and eventually form a gravitationally bound binary at pc or sub-pc scales \citep[e.g.,][]{DeRosa2020QP, Pfeifle2025ApJS}. Dual quasars are therefore considered as natural progenitors of merging SMBHs, and serve as essential and important laboratories for understanding the accretion, merger, and growth history of SMBHs \citep{Kauffmann2000, DiMatteo2005Nature}.
    
    Moreover, the inspiral and eventual coalescence of binary SMBHs are expected to generate low-frequency gravitational waves that dominate the stochastic background detected by pulsar timing arrays \citep{Chen2019MNRAS, Padmanabhan2024, 2024EPTA_Collaboration}, and will also be the prime targets for future space-based detectors such as LISA \citep{LISA2023}.

    However, a few hundred dual quasars have been confirmed to day \citep{Pfeifle2025ApJS, Tang2026MNRAS, Yue2026ApJ}. 
    The search for more such systems is hindered not only by the small angular separations between system members, which challenge the resolution of ground-based imaging and spectroscopy \citep{Liu2011ApJ, Hennawi2006}, but also by contamination in the candidate selection. Two major contaminants are gravitationally lensed quasars, where multiple images of a single background quasar produced by a foreground lens galaxy can closely mimic a genuine dual system \citep{Inada2012, Lemon2019, Yue2023, Dux2024}, and projected quasars, where two physically unrelated quasars at different redshifts appear close on the sky due to chance alignment. Nevertheless, ongoing and forthcoming wide-area spectroscopic and imaging surveys (e.g., SDSS or those of LAMOST, DESI, Euclid, LSST, and CSST) will dramatically increase the parent quasar sample and provide opportunities to increase the number of dual quasars \citep{Wu2010MN, Shi2014RAA, Silverman2020ApJ, Liu2010ApJa, Tang2021ApJ, Wang2025, DingXH2025}.


    We carried out a series of works to search for dual quasars based on Gaia data. \textsc{Paper I} describes the method and construction of a catalog of more than 4,000 quasar pair candidates, termed as MGQPC \citep{Qihang2025MGQPC}. The pair candidates are isolated by searching for Gaia sources with zero proper motions and parallaxes near known quasars in the Million Quasars v8 Catalog \citep{Flesch2023}.
    \textsc{Paper II} developed a machine-learning photo-$z$ framework based on \textsc{CatBoost} point estimates and \textsc{FlexZBoost} PDFs, which was used to prioritize 185 high-probability quasar pair candidates from the MGQPC catalogue for spectroscopic follow-up \citep{ZhuXY2026paperII}.
    A subset of MGQPCs has been confirmed as dual, lensed, or projected quasar systems via dedicated spectroscopic follow-up with the BFOSC/Xinglong 2.16 m, YFOSC/Lijiang 2.4 m and DBSP/Palomar 200-inch telescope (\citep{chen2026searchqp}).

    This paper, the third in our series, presents the spectroscopic confirmations of some MGQPC candidates primarily using the Dark Energy Spectroscopic Instrument Data Release 1 (DESI DR1) data. We report 17 dual quasars and 143 projected quasars, and present their finding charts and spectra. This paper is organized as follows. 
    Section~\ref{sec:data} describes the data sources, the selection methodology, and presents spectra and images. Notes on individual systems are provided in Section~\ref{sec:systems}, followed by a discussion of the overall sample in Section~\ref{discussion}. We give the conclusion in Section~\ref{sec:sum}.
    We adopt a flat $\Lambda$ cold dark matter ($\Lambda$CDM) cosmology throughout, with H$_0 = 70\ \mathrm{km\,s^{-1}\,Mpc^{-1}}$ and $\Omega_{\mathrm{m}} = 0.3$.

    \section{Data, Method, and Results} \label{sec:data}
    We used the spectra in the Spectra Analysis and Retrievable Catalog Lab (SPARCL) from the Astro Data Lab\footnote{\url{https://datalab.noirlab.edu/index.php}}, which contains spectra from SDSS DR16 (including BOSS DR16), DESI Early Data Release (DESI EDR), and the recently released DESI DR1\footnote{\url{https://data.desi.lbl.gov/doc/}} \citep{DESI2025dr1}. The DESI Legacy Imaging Surveys (DESI-LS) DR9 cutout images were utilized to display the paired and projected quasars reported in this work.

    \subsection{Data}
    The Dark Energy Spectroscopic Instrument (DESI) is mounted on the Mayall 4-meter telescope at Kitt Peak. It uses 5\,000 robotic fibers and a $3.2^{\circ}$ diameter field-of-view to acquire high-efficiency spectroscopic observations. Its targets are selected from the 14,000 deg$^2$ Legacy Imaging Surveys (DR9) based on optical and mid-infrared photometry \citep{DESI-LS2019}. On March 19, 2025, DESI released its first data release (DR1), which provides spectra for approximately 1.6 million quasars over more than 9000 deg$^2$ of sky.

    SDSS DR16\footnote{\url{https://www.sdss.org/dr16/}} is one of the results of SDSS-IV and provides optical spectra obtained with the 2.5 m Sloan Foundation Telescope at Apache Point Observatory \citep{Blanton2017, Ahumada2020}. The spectroscopic data set of DR16 delivers wide-area coverage and well-calibrated redshifts and classifications for large samples of galaxies and quasars \citep{Dawson2013, Ahumada2020}.

    LAMOST, also known as the Guoshoujing Telescope, is a quasi-meridian reflecting Schmidt telescope located at Xinglong Observatory, China. It has an effective aperture of 3.6--4.9 m and a wide field of view of about 5$^\circ$ in diameter. Its focal plane is equipped with 4000 fibers, each with a diameter of 3.3$\arcsec$, feeding 16 spectrographs and enabling highly efficient large-scale spectroscopic surveys \citep{Lyu2026ApJS}. The LAMOST quasar survey has provided a large number of spectroscopically confirmed quasars over its regular survey operations. In this work, we also use the LAMOST quasar sample as the spectroscopic dataset to confirm dual quasars.
    
    The MGQPC sample comprises 4,112 candidate quasar pairs. Each pair includes a known quasar from the MQC (hereafter, \texttt{member\_A}) and a candidate quasar (hereafter, \texttt{member\_B}) isolated by the astrometric method. The projected distances between them are less than 100 kpc. The median member separation of them is 8.81$\arcsec$, the Gaia G-band median magnitude is 20.52, and the median redshift is 1.61.
 
    \subsection{Method} 
    
    We first cross-matched the \texttt{member\_B} sources against the public spectroscopic archives via Astro Data Lab and the SPARCL database, yielding 424 matches, including 241 quasars, 111 stars, and 72 galaxies. Among the 241 new quasars, 49 were reported by \citet{Jing2026}, who identified more than 1,000 quasar pairs or candidates by self-matching the DESI DR1 quasar sample (the \texttt{member\_A} quasars in the 49 pairs were reobserved by the DESI project), and were removed from our sample. This leaves 192 new quasars. Then, a query of the latest Strong Gravitational Lensing Database (SLED)\footnote{\url{https://sled.amnh.org/}} removed 22 systems that had previously been identified as 12 confirmed lenses, 6 lens candidates, 2 quasar pairs, 1 projected quasar, and 1 quasar+star.
    After that, we searched the remaining 170 systems in the NASA/IPAC Extragalactic Database (NED\footnote{\url{https://ned.ipac.caltech.edu/}}) and excluded one quasar pair system that was reported most recently. 
    
    To avoid cases where the automated template fitting failed for spectra with low signal-to-noise ratios and led to redshift calibration errors, we visually inspected the spectra of these 169 pairs. As a result, 6 redshifts were modified, and 1 was discarded since the spectral quality was too poor to provide a reliable redshift. This leaves 168 quasars.

    Following \citet{Hennawi2010deltaV}, a radial velocity difference of $\lvert \Delta v_r \rvert \leqslant 2000\ \mathrm{km\,s^{-1}}$ was adopted as the criterion for dual quasars with broad emission lines at high redshift. The $\lvert \Delta v_r \rvert$ was calculated with the following formula \citep{hogg2000distancecosmos}:
    \begin{equation}
    \lvert \Delta v_r \rvert = c\cdot\frac{\lvert z_A - z_B \rvert}{1 + (z_A + z_B)/2},
    \end{equation}
    \label{Eq1}where $z_A$ and $z_B$ are the redshifts of the two quasar members. 
    With this criterion, 22 new quasars were found to form dual systems with the known \texttt{member\_A} quasars, including 5 candidate dual quasars with their \texttt{member\_A} quasars having only photometric redshifts. The remaining 143 new quasars form projected systems with their known \texttt{member\_A} quasars. In addition, three candidate pairs were identified as unresolved single sources in the original data and were therefore excluded from the final sample.

    \subsection{Results}
    The images and spectra of the 17 new dual quasars are presented in Figures~\ref{DQ1}. The range of redshifts is from 0.573 to 2.758, with transverse distances of 12.157$-$96.855 kpc. We will provide a detailed description of these systems in the next section.
    Tables~\ref{DQ_parameters}, \ref{5photoz}, \ref{PQ_parameters} also list the basic parameters for the 17 dual quasars, 5 candidate dual quasars, and 143 projected quasars, respectively. It should be emphasized that the five systems with only photometric redshift lack sufficient redshift precision for a robust determination of their line-of-sight(LOS) velocity difference. We therefore treat them as candidate systems throughout this work.

    The \texttt{member\_B} quasar candidates in the MGQPC sample were isolated based on the astrometric method. The purity of the quasar candidate sample is 56.8\%, based on 241 quasars identified among 424 spectra. Among the 241 quasars, about 27.8\% form dual quasars with the known \texttt{member\_A} quasars, including the quasar pairs reported in \citet{Jing2026}, while the remainings are projected ones.
    
    \begin{figure*}
    \centering
    \includegraphics[width=17cm]{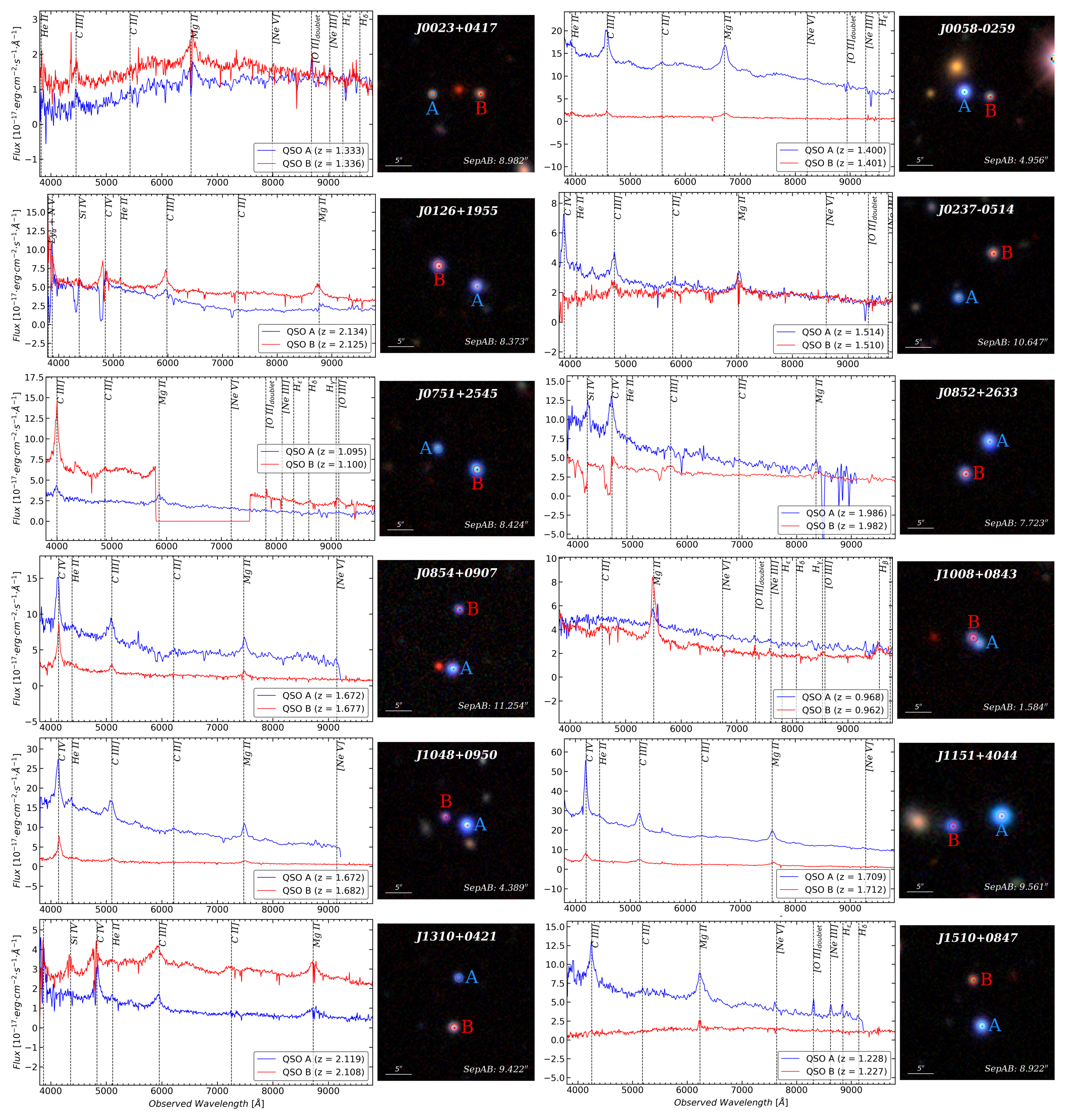}
    \caption{Spectra and DESI-LS DR10 pseudo-color cutout images of the 17 dual quasars. 
    For each system, the blue and red spectra are of the blue (\texttt{member\_A}) and red (\texttt{member\_B}) objects in the images, respectively. The black dashed vertical lines mark the wavelengths of the major emission lines at the redshift of \texttt{member\_A}. The angular separation between the two quasars is given as SepAB in the lower-right corner of each image. 
    Each image covers a $30^{\prime\prime}\times30^{\prime\prime}$ field of view, with north to the up and east to the left, and is constructed from the DESI-LS DR10 $g$-, $r$-, and $z$-band imaging data using HumVI \citep{Marshall2015SWHumVI, Marshall2016HumVI}.}
    \label{DQ1}
    \end{figure*}

    \begin{figure*}
    \addtocounter{figure}{-1} 
    \centering
    \includegraphics[width=17cm]{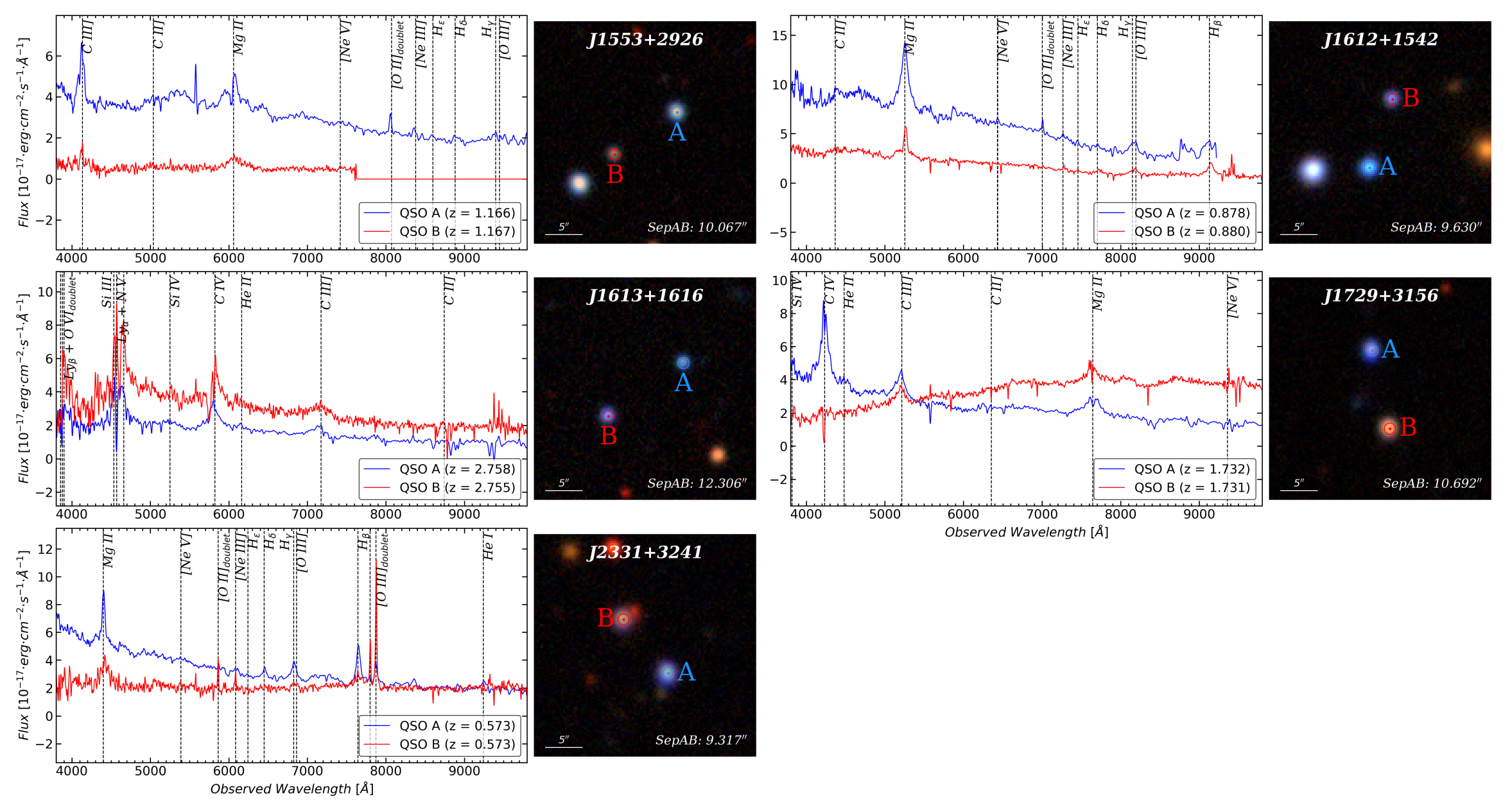}
    \caption{- \textit{continued}}
    \end{figure*}
    
\section{Notes on individual systems} \label{sec:systems}
    In this section, we present the images and spectra for our newly confirmed pairs and provide brief notes on individual systems, and highlight two close projected pairs.
    
    \subsection{Dual quasars}
    
    \subsubsection{J0023+0417} 
    The two members are at z $\sim$ 1.334 and are separated by 8.982$\arcsec$. Their spectra are very similar, with absorptions in the Mg II emission in both. The image shows a red object between the two quasars. These characteristics are also compatible with a lensing scenario. We therefore classify this system as a nearly identical quasar (NIQ) \citep{Anguita2018NIQ, Lemon2018GLQG-II24LQ, Lemon2020STRIDES10LQ+10QP, Lemon2023}, and retain it for future analysis and follow-up observations.

    \subsubsection{J0058+0259}
    The two quasars are at z$\sim$1.401, with continua and emission lines markedly different in strength. They are separated by 4.956$\arcsec$. No plausible foreground deflector is visible in the current image. We therefore classify this system as a dual quasar.
    
    \subsubsection{J0126+1955} 
    The blue spectrum shows broad and strong blueshifted Ly~$\alpha$, Si IV, and C IV absorption troughs, suggestive of a broad absorption line (BAL; \citealp{Weymann1991ApJ}) quasar. No similar BAL feature is observed in the red spectrum. We designate this system as a distinct dual quasar. The separation of the two memebers is 8.373$\arcsec$, which corresponds to a transverse distance of 69.381 kpc at z $\sim$ 2.130.

    \subsubsection{J0237-0514}
    Both members exhibit characteristic quasar features at nearly identical redshifts (z$\sim$1.512), yet there are noticeable differences in the strength of C IV and continuum slope shortward of $\sim$7000~\AA. The presence of an absorption doublet profile at the line center in \texttt{member\_B} and its absence in \texttt{member\_A}, provide strong evidence that the two members are intrinsically different, favoring a dual quasar scenario.
    
    \subsubsection{J0751+2545} 
    The system is located at z $\sim$ 1.098, with a separation of 8.424\arcsec and a transverse distance of 68.598 kpc. 
    The continuum of \texttt{member\_B} is partly missing, yet the marked difference in the strength of the C III] emission line suggests that the two members are physically distinct sources.
    
    \subsubsection{J0852+2633} 
     This system consists of two quasar members at z $\sim$ 1.984, separated by 7.723\arcsec and 68.649 kpc. The BAL features shortward of Si IV and C IV lines in the red spectrum suggest that the two members are intrinsically distinct, supporting a dual quasar interpretation for the system.

    \subsubsection{J0854+0907} 
    The spectra indicate that both members are quasars at \(z\sim1.674\). The C III] emission lines have substantially different widths in the two members, with measured FWHM values of \(8768\) and \(3437~{\rm km~s^{-1}}\), respectively. No plausible foreground deflector is seen in the image. Therefore, we classify this system as a distinct dual quasar, with a transverse distance of 95.28 kpc and a separation of 11.254$\arcsec$.

    \subsubsection{J1008+0843} 
    The continuum flux of \texttt{member\_A} is consistently higher than that of \texttt{member\_B} over the full wavelength range, while \texttt{member\_B} exhibits markedly stronger Mg II emission, indicating that the two members are physically distinct. The separation is 11.255$\arcsec$ at z$\sim$0.965, and the transverse distance is 64.45 kpc.
    
    \subsubsection{J1048+0950}
    The spectra of the two members are consistent with quasars at z$\sim$1.677, with a separation of 4.389$\arcsec$ and a transverse distance of 36.9 kpc. Significant differences in strength and profile (see left shoulders of the C IV and C III] lines in the blue spectrum) can be seen in the emission lines. Together with the absence of deflators between the members in the image, we favor a dual quasar classification. 
    
    \subsubsection{J1151+4044} 
    The spectra show two quasars at \(z=1.710\). 
    The two components exhibit different broad-line profiles, most clearly in C~IV, for which we measure FWHM values of \(5141.7\) and \(7555.5~{\rm km~s^{-1}}\) for components A and B, respectively. 
    Together with the absence of a plausible lensing galaxy in the image, these spectral differences favour a dual-quasar interpretation. 
    The angular separation is \(9.561\arcsec\), corresponding to a projected transverse distance of \(80.42~{\rm kpc}\).

    \subsubsection{J1310+0421}
    The spectra reveal two quasars at $z\sim2.113$. Both spectra show clear self-absorptions in the C IV and Mg II emissions. Their continua have similar slopes longward of C III] lines but very different slopes shortward of C III] lines, indicating different origins of the two continua. Together with the absence of any obvious lens in the image, we classify the system as a dual quasar, with a transverse distance of 78.27 kpc and an angular separation of 9.422$\arcsec$.
    
    
    \subsubsection{J1510+0847} 
    The system consists of two quasars at z$\sim$ 1.227. The two spectra show striking differences in both continuum and emission-line strength, indicating that the two members are intrinsically different quasars and form a dual system with member separations of 8.922$\arcsec$ and 74.08 kpc.

    \subsubsection{J1553+2926} 
    The spectra show two quasars at z=1.166, separated by 10.067$\arcsec$ (corresponding to 82.745 kpc). The two spectra have very different strength and profile in C III] and Mg II emission lines. Therefore, we classify this system as a dual quasar.

    \subsubsection{J1612+1542} 
    \texttt{member\_A} shows a substantially stronger and broader Mg~II profile than component B, with measured Mg~II FWHM values of \(4911.6\) and \(3247.3~{\rm km~s^{-1}}\), respectively. 
    We therefore classify this system as a dual quasar at $z\sim0.879$, with a separation of 9.630$\arcsec$ and a transverse distance of 74.36 kpc.

    \subsubsection{J1613+1611} 
    \texttt{member\_B} shows absorption on the blue wing of the C IV emission line. However, \texttt{member\_A} does not show comparable absorption at the same wavelength. Therefore, we consider that this is a dual quasar, with a separation of 12.306$\arcsec$, and a 96.845 kpc in transverse distance at $z \sim$ 2.756

    \subsubsection{J1729+3156}
    The spectra confirm that both members are quasars at $z \sim 1.731$ (separation =10.692$\arcsec$, $r_p=90.14$ kpc). \texttt{member\_A} and \texttt{member\_B} exhibit markedly different C IV emission line, while \texttt{member\_B} shows a redder continuum. We therefore classify the system as a dual quasar.

    \subsubsection{J2331+3241} 
    The blue (\texttt{member\_A}) spectrum shows moderately broad Mg II and Balmer lines, strong and broad Fe II lines, and weak [O III] doublet, which are typical features of narrow-line Seyfert 1 galaxies. However, the red (\texttt{member\_B}) spectrum has broad Mg II and H~$\beta$ lines and strong [O III] lines. These spectral differences indicate the different identities of two objects and classify them as a dual quasar system. The member separation is 9.317\arcsec and the transverse distance is 61.05\,kpc.

\subsection{Projected quasars}
\begin{figure}
    \centering
    \includegraphics[width=9cm]{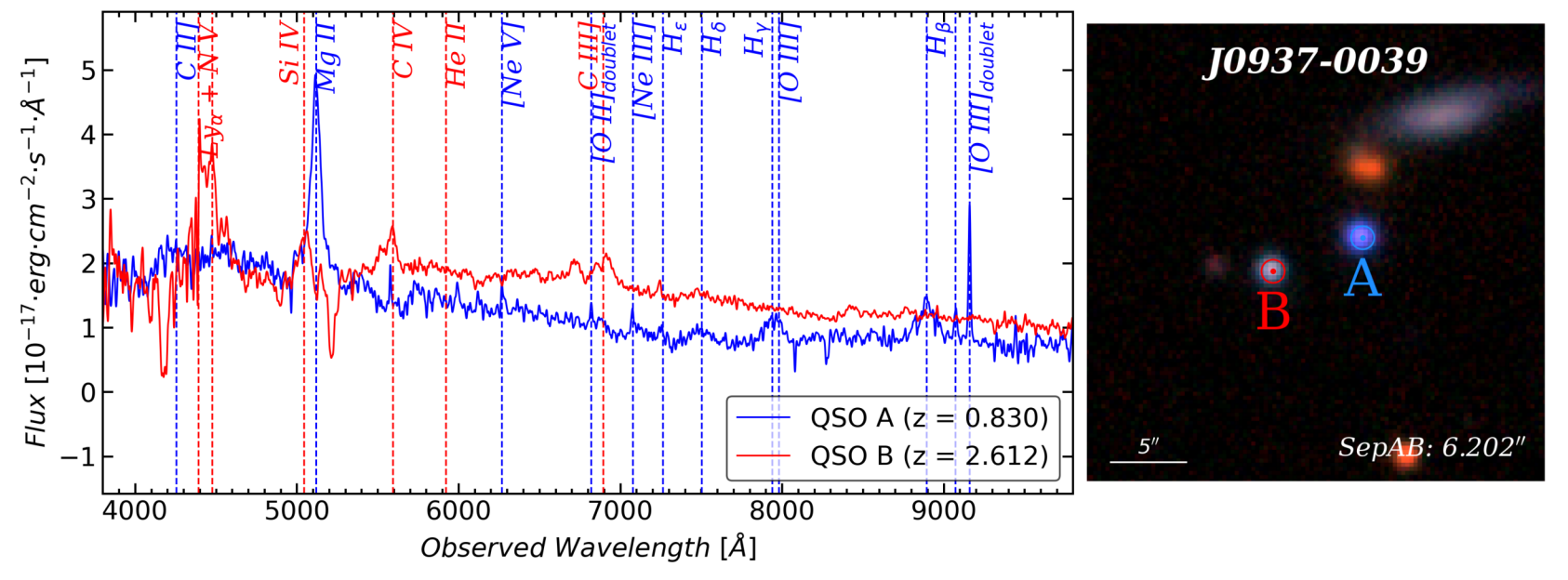}
    \caption{The spectra and images of projected quasar J0909+1509 and J0937-0039. The symbles and colors are the same as in Fig~\ref{DQ1}. }
    \label{PQJ0937-0039}
    \end{figure}

    In addition to the 17 dual quasars, we also identified 143 new projected quasars. Their redshifts span 0.324 to 4.030, with projected distances of 17.467–99.946 kpc (Table~\ref{PQ_parameters}). Although most projected quasars have r$_p \gtrsim$ 50 kpc, a subset has relatively small projected distances. J0937$-$0039 is one such system, as shown in Figure~\ref{PQJ0937-0039}. Such foreground–background quasar configurations can provide useful sightlines for studying gas associated with the foreground quasar host galaxy and its circumgalactic environment through absorption features imprinted on the background spectrum \citep{Hada2025}.

    However, such absorption features cannot always be unambiguously attributed to the environment of the foreground quasar. They may instead arise from intrinsic absorption associated with the background quasar itself, including absorption produced by quasar-driven outflows \citep{FilizAk2013ApJ, Capellupo2013MNRAS, Hamann2019MNRAS}, or from unrelated intervening absorbers along the line of sight \citep{Churchill2013ApJ}. Therefore, the origin of absorption should be interpreted with caution.



    \begin{table*}
    \caption{The basic parameters of 17 dual quasars.}
    \label{DQ_parameters}
    \centering
    \begin{tabular}{lrrrrrrr}
    \hline\hline
    \multicolumn{1}{c}{Name$^{\,(1)}$} &
    \multicolumn{1}{c}{R.A.$^{\,(2)}$}          &
    \multicolumn{1}{c}{Dec$^{\,(3)}$}         &
    \multicolumn{1}{c}{$G$$^{\,(4)}$}           &
    \multicolumn{1}{c}{Redshift$^{\,(5)}$}    &
    \multicolumn{1}{c}{SepAB$^{\,(6)}$}         &
    \multicolumn{1}{c}{$\overline{r_p}$$^{\,(7)}$}       &
    \multicolumn{1}{c}{$\lvert \Delta v_r \rvert$$^{\,(8)}$}\\
    & 
    \multicolumn{1}{c}{(J2000.0)}   &
    \multicolumn{1}{c}{(J2000.0)}   &
    \multicolumn{1}{c}{(mag)}       &
    & 
    \multicolumn{1}{c}{($\arcsec$)} &
    \multicolumn{1}{c}{(kpc)}       &
    \multicolumn{1}{c}{(km\,s$^{-1}$)}\\
    \hline
    J0023+0417A & 5.909058  &  4.283466  & 21.01 & 1.33300 & \multirow{2}{*}{8.982}  & \multirow{2}{*}{76.742} & \multirow{2}{*}{385.589} \\
    J0023+0417B & 5.906513  &  4.283475  & 20.80 & 1.33608 &  &  &  \\
    \hline
    J0058$-$0259A & 14.562506  & $-$2.992344  & 18.56 & 1.39957 & \multirow{2}{*}{4.956} & \multirow{2}{*}{42.673} & \multirow{2}{*}{175.637} \\
    J0058$-$0259B & 14.561123  & $-$2.992604  & 20.68 & 1.40097 &   &  &  \\
    \hline
    J0126+1955A & 21.728592  & 19.916194  & 19.62 & 2.13400 & \multirow{2}{*}{8.373}  & \multirow{2}{*}{69.402} & \multirow{2}{*}{853.947} \\
    J0126+1955B & 21.730773  & 19.917281  & 19.59 & 2.12509 &  &  & \\
    \hline
    J0237$-$0514A & 39.473810 & $-$5.248319  & 20.30 & 1.51359 & \multirow{2}{*}{10.647} & \multirow{2}{*}{89.906} & \multirow{2}{*}{428.776} \\
    J0237$-$0514B & 39.471973 & $-$5.246005  & 20.32 & 1.51000 &  &  & \\
    \hline
    J0751+2545A & 117.918240 & 25.765423  & 20.43 & 1.09539 & \multirow{2}{*}{8.424}  & \multirow{2}{*}{68.624} & \multirow{2}{*}{616.877} \\
    J0751+2545B & 117.915950 & 25.764331  & 19.32 & 1.09971 &  &  & \\
    \hline
    
    J0852+2633A & 133.205260 &  26.563328  & 18.89 & 1.98584 & \multirow{2}{*}{7.723} & \multirow{2}{*}{64.457} & \multirow{2}{*}{345.591} \\
    J0852+2633B & 133.206690 &  26.561616  & 19.68 & 1.98240 &  &  & \\
    \hline
    J0854+0907A & 133.619350 &  9.125723  & 18.96 & 1.67193 & \multirow{2}{*}{11.255} & \multirow{2}{*}{95.285} & \multirow{2}{*}{563.315} \\
    J0854+0907B & 133.619018 &  9.128831  & 20.54 & 1.67695 &  &  & \\
    \hline
    J1008+0843A & 152.003377 & 8.727811 & 19.96 & 0.96818 & \multirow{2}{*}{1.584} & \multirow{2}{*}{12.157} & \multirow{2}{*}{969.174} \\
    J1008+0843B & 152.003705 & 8.728086 & 20.40 & 0.96183 &  &  & \\
    \hline
    J1048+0950A & 162.192540 &  9.836522  & 18.41 & 1.67163 & \multirow{2}{*}{4.389}  & \multirow{2}{*}{36.908} & \multirow{2}{*}{1137.629}\\
    J1048+0950B & 162.193690 &  9.836950  & 20.46 & 1.68178 &  &  &  \\
    \hline
    J1151+4044A & 177.761480 & 40.737962  & 17.76 & 1.70925 & \multirow{2}{*}{9.561}  & \multirow{2}{*}{80.420} & \multirow{2}{*}{348.960} \\
    J1151+4044B & 177.764888 & 40.737417  & 19.68 & 1.71240 & &  &  \\
    \hline
    J1310+0421A & 197.552398 & 4.353193  & 20.84 & 2.11892 & \multirow{2}{*}{9.422} & \multirow{2}{*}{78.301} & \multirow{2}{*}{1010.183}  \\
    J1310+0421B & 197.552638 & 4.350588  & 20.11 & 2.10843 &  &  &  \\
    \hline
    J1510+0847A & 227.710990 &  8.787205  & 19.08 & 1.22821 & \multirow{2}{*}{8.922}  & \multirow{2}{*}{74.075} & \multirow{2}{*}{218.547} \\
    J1510+0847B & 227.711455 &  8.789636  & 20.81 & 1.22658 &  &  &  \\
    \hline
    J1553+2926A & 238.482470 & 29.443562  & 19.60 & 1.16557 & \multirow{2}{*}{10.067} & \multirow{2}{*}{82.755} & \multirow{2}{*}{232.061} \\
    J1553+2926B & 238.485123 & 29.442009  & 20.68 & 1.16724 &  &  &  \\
    \hline
    J1612+1542A & 243.064550 & 15.713796  & 19.81 & 0.87783 & \multirow{2}{*}{9.630}  & \multirow{2}{*}{74.389} & \multirow{2}{*}{417.023} \\
    J1612+1542B & 243.063666 & 15.716327  & 20.34 & 0.88045 &  &  &  \\
    \hline
    J1613+1616A & 243.326580 & 16.278684  & 20.76 & 2.75837 & \multirow{2}{*}{12.306} & \multirow{2}{*}{96.860} & \multirow{2}{*}{261.391} \\
    J1613+1616B & 243.329485 & 16.276718  & 20.51 & 2.75510 &  &  &  \\
    \hline
    J1729+3156A & 262.348950 & 31.941437  & 19.96 & 1.73221 & \multirow{2}{*}{10.692} & \multirow{2}{*}{90.143} & \multirow{2}{*}{124.592} \\
    J1729+3156B & 262.348187 & 31.938549  & 19.42 & 1.73107 &  &  &  \\
    \hline
    J2331+3241A & 352.751530 & 32.694258  & 20.02 & 0.57258 & \multirow{2}{*}{9.317}  & \multirow{2}{*}{61.035} & \multirow{2}{*}{123.629} \\
    J2331+3241B & 352.753503 & 32.696248  & 20.76 & 0.57323 &  &  &  \\
    \hline
    \end{tabular}
    \tablefoot{Columns: (1) System name, member tag (A or B) used to distinguish the two members of each system ;(2–3) Right ascension and declination (degrees); (4) $G$-band magnitude; (5) Spectroscopic redshifts; (6-8) Angular separation, transverse distance at the mean pair redshift, and the LOS velocity difference.}
    \end{table*}

    \begin{table*}[h!]
        \caption{The primary parameters of five dual quasar candidates.}
        \label{5photoz}
        \centering
        \begin{tabular}{lrrrll}
        \hline\hline
        \multicolumn{1}{c}{Name$^{\,(1)}$} &
        \multicolumn{1}{c}{R.A.$^{\,(2)}$}          &
        \multicolumn{1}{c}{Dec$^{\,(3)}$}         &
        \multicolumn{1}{c}{$G$$^{\,(4)}$}           &
        \multicolumn{1}{c}{Redshift$^{\,(5)}$}    &
        \multicolumn{1}{c}{SepAB$^{\,(6)}$} \\
        &
        \multicolumn{1}{c}{(J2000.0)}   &
        \multicolumn{1}{c}{(J2000.0)}   &
        \multicolumn{1}{c}{(mag)}       &
        \multicolumn{1}{c}{($\arcsec$)} \\
        \hline
        J0353$-$1145A & 58.296961  & $-$11.760595 & 21.00 & 1.00 & \multirow{2}{*}{6.856} \\
        J0353$-$1145B & 58.296627  & $-$11.758721 & 20.71 & 1.51470 &  \\
        \hline
        J1224+0123A & 186.202546 & 1.395552 & 19.18 & 1.38 & \multirow{2}{*}{2.419}\\
        J1224+0123B & 186.203153 & 1.395789 & 19.76 &1.38564   \\
        \hline
        J1710+6411A & 257.689830 & 64.198101 & 19.67 & 2.20 & \multirow{2}{*}{6.109} \\
        J1710+6411B & 257.692948 & 64.197075 & 19.97 & 0.77842 &  \\
        \hline
        J1816+4340A & 274.053125 & 43.669347 & 20.43 & 2.55 & \multirow{2}{*}{7.267} \\
        J1816+4340B & 274.054935 & 43.667811 & 20.26 & 1.22549 &  \\
        \hline
        J1836+5235A & 279.069146 & 52.599571 & 19.07 & 1.81 & \multirow{2}{*}{2.199} \\
        J1836+5235B & 279.069506 & 52.599010 & 19.13 & 1.81970 &  \\
         \hline
        \end{tabular}
        \tablefoot{Columns: (1) System name, member tag (A or B) used to distinguish the two members of each system ;(2–3) Right ascension and declination (degrees); (4) $G$-band magnitude; (5) Photo/Spectroscopic redshifts; (6) Angular separation.}
        \end{table*}
        
\section{Discussion} 
\label{discussion}
    \subsection{Dual quasar}
\begin{figure}
    \centering
    \includegraphics[width=9cm]{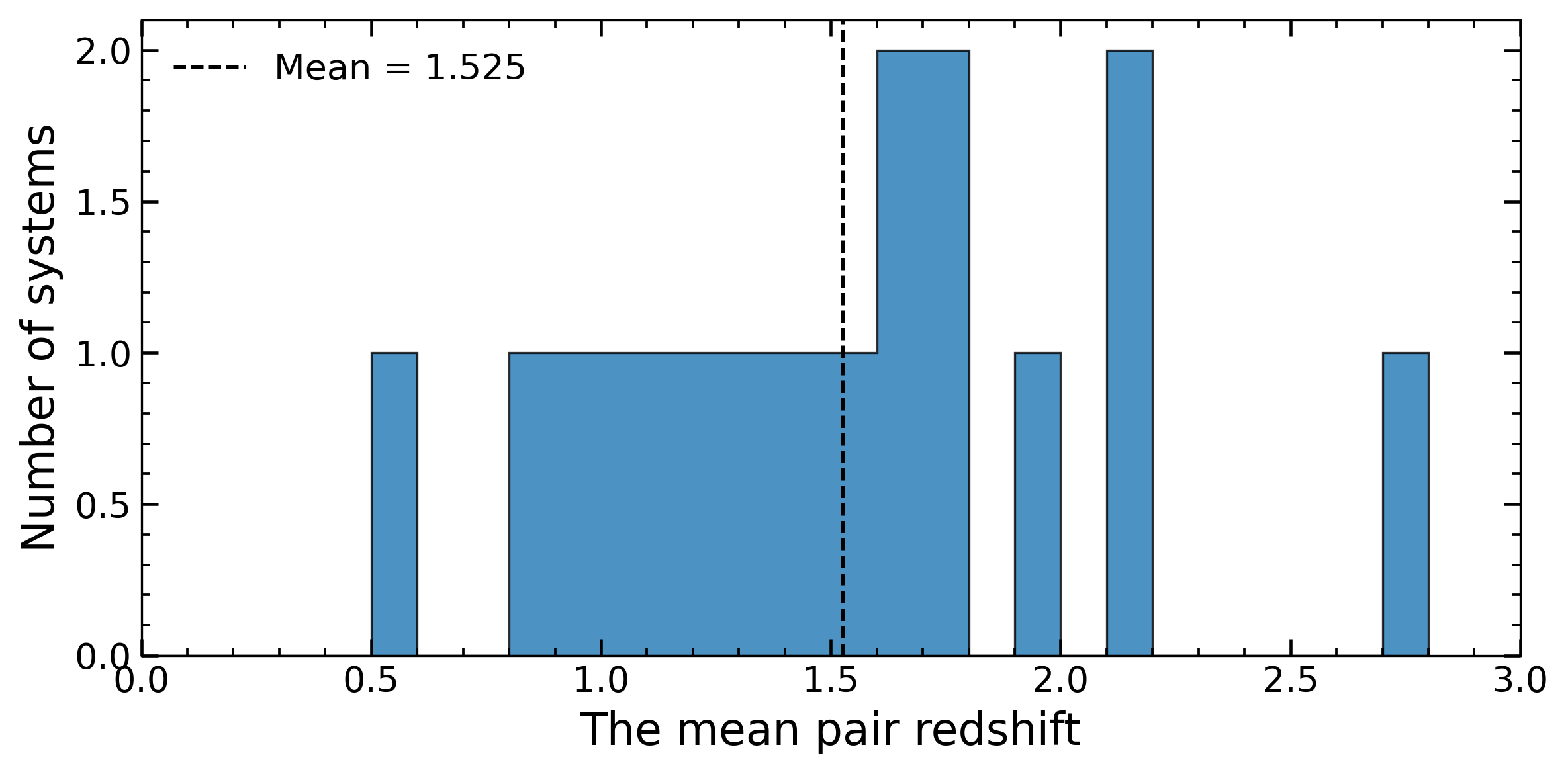}
    \caption{Distribution of the mean redshift of dual quasars. The dashed vertical line marks the sample mean, $\langle z\rangle$= 1.525.}
    \label{z_distribution}
    \end{figure}

    The pairs have a mean redshift of $\langle z\rangle$ = 1.525 and extend to a maximum of z = 2.758, with nine systems at z $\ge$ 1.5. Figure \ref{z_distribution} shows the distribution of the mean pair redshifts. 

    Our sample comprises dual quasars with separations of \(\sim4\arcsec\) to \(12\arcsec\), which can be spatially resolved in most wide-field imaging surveys. Over this separation range, our selection is less affected by several issues that commonly complicate closer-pair ($<$ 3\arcsec) searches: they may be unresolvable and missed in seeing-limited data \citep{Liu2011ApJ, Fu2011a, Comerford2012ApJ}, while the spectroscopic identification may suffer from source confusion due to fibre spillover effects \citep{Husemann2018, Pfeifle2023ApJ} or from extended narrow-line regions that may mimic dual-nucleus signatures \citep{H.Fu2012ApJ, H.Fu2018ApJ, Keel2019MNRAS}. 
    It is worth noting that the spectra used for our sample was not obtained contemporaneously. Since each AGN can exhibit intrinsically variable, the observed differences between the two components may reflect both their genuine source differences and epoch-dependent variations in the continuum shape, emission-line equivalent widths, and flux calibration (e.g., \citealt{MacLeod2010, Schmidt2010, Salvato2011}). We therefore treat mild spectral differences with caution and do not regard them as decisive evidence for a dual quasar classification. 
    The classification is instead based on a combination of diagnostics, including $\lvert \Delta v_r \rvert$, emission-line profile differences, BAL presence or absence, source morphology, and the visibility of a plausible foreground deflector. Thus, the identification of dual quasars in our sample is based on a holistic assessment of the available spectroscopic and imaging information.

    One remaining concern is whether a subset could be strong lensing systems rather than dual quasars. In several systems, the spectra of the two members are not fully matched in the continuum shape and in the relative strengths and detailed profiles of prominent emission lines (e.g., C IV and Mg II). Under a strong-lensing hypothesis, such spectral differences could arise from intrinsic quasar variability combined with the lensing time delay, because the lensed images are observed at different source epochs. potentially producing modest continuum-level mismatches \citep{Eigenbrod2008A&A, MacLeod2010, Millon2020, Shajib2020}. Additional differences may arise from differential extinction along the distinct sightlines through the lens galaxy, as well as from microlensing by stars in the lens galaxy, which primarily affects the compact emitting region \citep{Wambsganss2006, Mosquera2013ApJ, Jimenez2015ApJ} and can therefore modify the apparent equivalent widths \citep{Falco1999ApJ, Eliasdottir2006ApJS, Sluse2012A&A}. However, for the wide-separation subset ($\gtrsim4\arcsec$, and in case even up to $\sim9\arcsec$), producing such wide separations via lensing generally requires a relatively massive deflector, often associated with a group/cluster-scale potential \citep{Oguri2009MNRAS,Oguri2010MNRAS}. Since our available images do not reveal obvious deflector galaxy between (or close to) the members, the combined spectral and image information disfavors a simple two-image lens interpretation for these systems and supports their classification as dual quasars.

\subsection{Dual quasar fraction}
\begin{figure}[htpb]
    \centering
    \includegraphics[width=9cm]{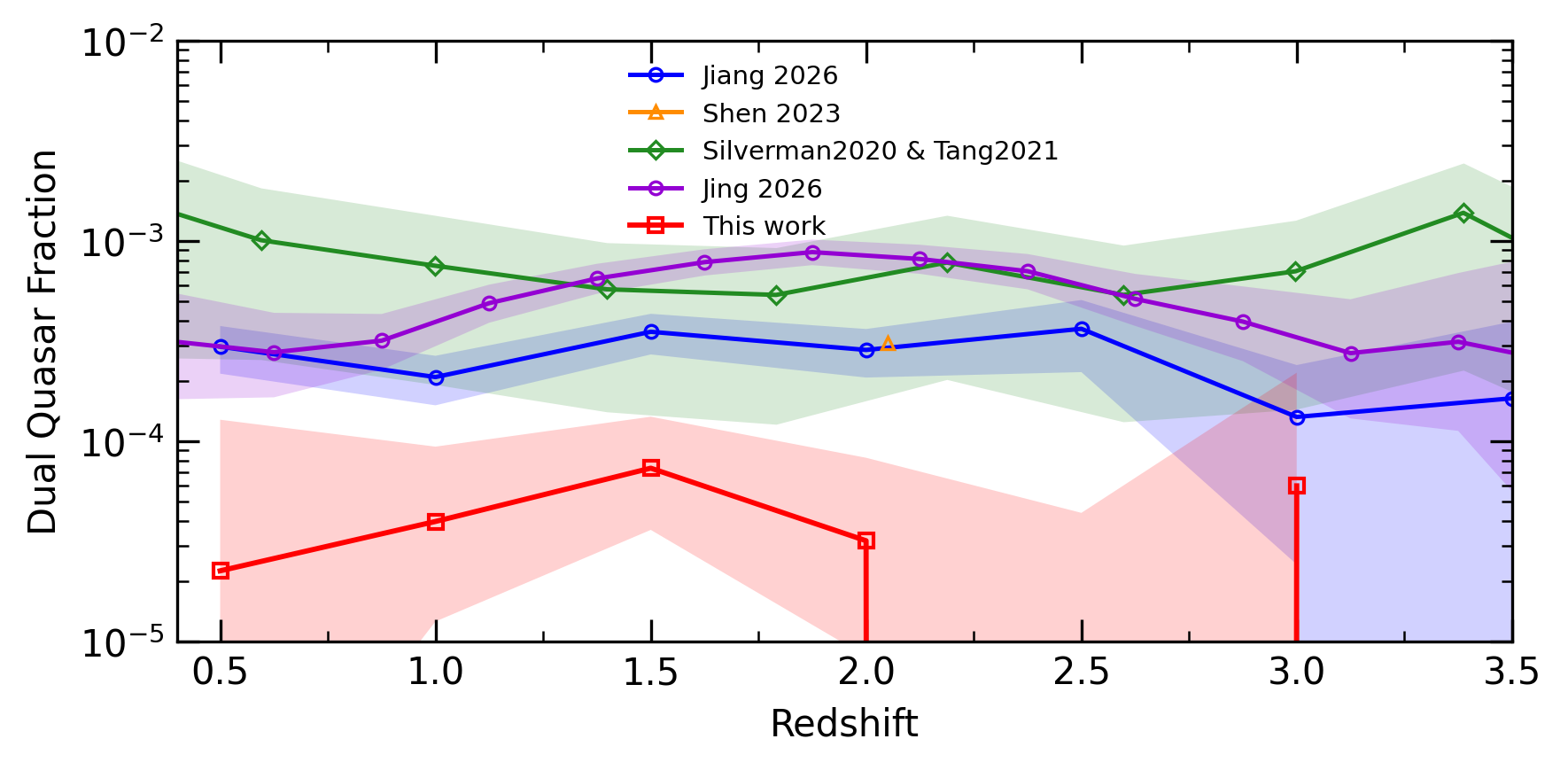}
    \caption{Dual quasar fraction as a function of redshift. 
    The red squares represent the fraction measured in this work, while the blue circles, orange triangles, green diamonds, and purple circles show the results from \citet{Jiang2026ApJ}, \citet{ShenY2023QPfraction}, \citet{Silverman2020ApJ,Tang2021ApJ}, and \citet{Jing2026}, respectively. 
    The dual quasar fractions are computed in fixed redshift bins of width $\Delta z = 0.75$.  
    The shaded regions denote the corresponding $2\sigma$ uncertainties.   
    Our fraction should be regarded as a lower limit to the intrinsic dual quasar fraction in the MGQPC sample, since it is based only on currently confirmed systems.}
    \label{fraction}
    \end{figure}

    Figure~\ref{fraction} shows the dual quasar fraction derived from the combined sample of 23 dual quasars, including six systems reported in \citep{chen2026searchqp}.
    The parent sample is based on the 582 725 quasars from \citep{Qihang2025MGQPC}, which were selected from MQCv8 with $z > 0.5$ and retained only when a Gaia optical counterpart was available. The fraction remains at the level of about $<10^{-4}$ over the redshift range $0.5 < z < 3.0$, with a slight enhancement around z about 1–2. At $z > 3$, the inferred fraction is highly uncertain due to the scarcity of sources in this redshift range.

    Compared with previous works that generally reported quasar pair fractions above 10$^{-4}$ at $z > 0.5$ \citep{Silverman2020ApJ, Tang2021ApJ, ShenY2023QPfraction, Jing2026, Jiang2026ApJ}, our fraction is generally lower.
    This is mainly because our sample is restricted to currently confirmed pairs, with many candidates still lacking spectroscopic observations. Therefore, it should be regarded as a current observed lower limit of our MGQPC rather than as an intrinsic dual quasar fraction.

    Nevertheless, the weak redshift dependence seen in our fraction, with only a mild enhancement around z about 1–2, is similar trend to that found for quasar pairs selected from DESI DR1 by \citep{Jing2026}. The enhancement coincides with the epoch of peak cosmic star formation and black-hole accretion activity. During this period, higher gas fractions and more frequent mergers may increase the probability of triggering both supermassive black holes in an interacting system \citep{Steinborn2016MNRAS, Capelo2017MNRAS, Rosas-Guevara2019MNRAS, Volonteri2022MNRAS, 2023MNRAS.522.1895C, Tang2026MNRAS}. However, Around $z=3$, the apparent slight increase in the fraction may be affected by small-number statistics, as this redshift bin contains only two dual quasar pairs and the number of parent quasars also decreases, making the inferred fraction more susceptible to stochastic fluctuations.

    It is worth noting that the fraction estimated here is subject to several selection effects. 
    Our optical selection is generally biased towards unobscured quasars, and may miss systems with obscured or optically faint companions \citep{Komossa2003, Satyapal2014, Ricci2017, Ricci2021, Capelo2017majormerge, Jing2026}, especially sources fainter than the Gaia limit of $G>21$ \citep{Qihang2025MGQPC}, leading to an incomplete census of dual quasars. 
    In addition, the current confirmed sample is still limited in size and is compiled from heterogeneous follow-up observations, so the inferred redshift dependence should be interpreted with caution. 
    We are continuing spectroscopic follow-up observations to enlarge the confirmed sample, and future larger, deeper, and more homogeneous observations, ideally combined with multi-wavelength diagnostics in the infrared, X-ray, and radio bands, will be required to obtain a more robust view of the evolution of the quasar pair fraction.
 
  \subsection{Interesting system: J0023+0417}
    Strongly lensed quasars are valuable tools for precision cosmology \citep{Courbin2018, Millon2020, Bonvin2017} and for studying the coevolution of supermassive black holes and their host galaxies through mass reconstruction \citep{Ding_2016, Millon2023}.
    Because lensed quasars often exhibit resemble dual quasars observationally, new dual quasars are sometimes serendipitously discovered during lens searches \citep{Lemon2019, Lemon2023, Dux2024, Yue2023}. Conversely, we anticipate that lensed quasars may be serendipitously identified during quasar pair confirmation.  
    
    Through visual inspection, a potential foreground lens is visible between the two members of J0023+0417 (see figure~\ref{DQ1}). The broad emission-line profiles of both members, including C III] and Mg II, are nearly similar, and both spectra show self-absorption features in the Mg II line. In addition, the redshift difference is only $\Delta z = 0.0031$. These characteristics strongly favor the interpretation that J0023+0417 is a gravitationally lensed quasar system.
    
\begin{figure}[htpb]
    \centering
    \includegraphics[width=9cm]{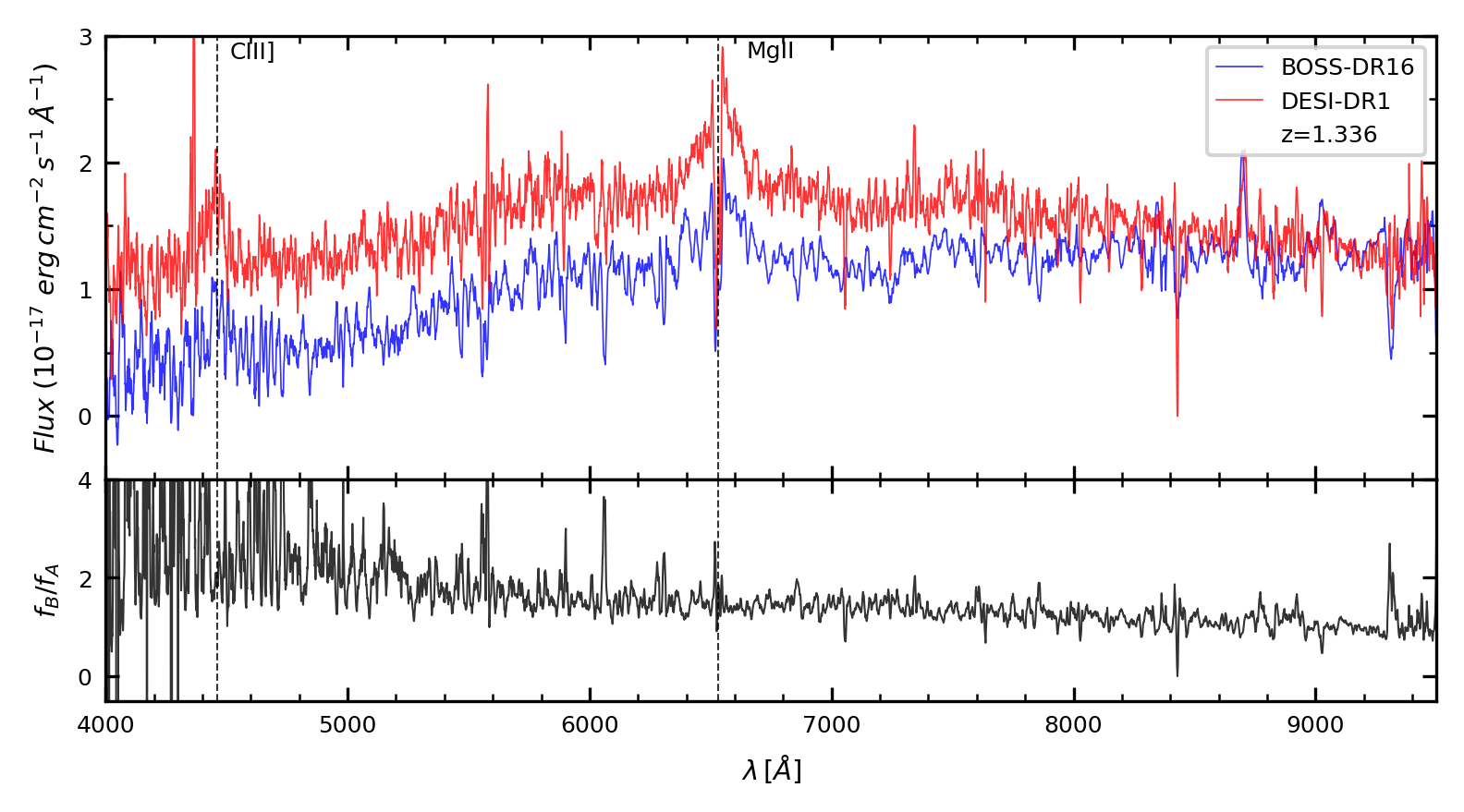}
    \caption{Top panel: Enlarged view of the spectra of the two quasar components in J0023+0417, corresponding to the spectra presented in Fig.~\ref{DQ1}. 
    The blue and red spectra represent J0023+0417A and J0023+0417B, obtained from BOSS DR16 and DESI DR1, respectively. 
    Bottom panel: Flux ratio between the two spectra, defined as $f_B/f_A$.}
    \label{fluxratio}
    \end{figure}

    Despite their striking similarity, subtle spectral differences between the two members are observed.
    First, the overall continuum of QSO A appears systematically brighter, especially at shorter wavelengths, producing a mild blue excess.
    This wavelength-dependent flux ratio is a characteristic signature of microlensing, in which stars in the lensing galaxy differentially magnify emission regions of different sizes, resulting in chromatic variation \citep{Courbin2002A&A, Eigenbrod2008A&A}. Second, the Mg II emission line in both spectra shows clear self-absorption features, while the absorption trough in QSO A is slightly deeper than that in QSO B.
    Such small differences in line depth and profile could also arise from microlensing effects, since the broad-line region (BLR) is only partially resolved by the lensing magnification pattern \citep{Sluse2012A&A}.
    Moreover, since the two spectra were obtained several years apart (BOSS DR16 2014–2018 and DESI DR1 2021–2022), intrinsic quasar variability must also be considered. Quasars are known to vary in the UV/optical continuum on timescales of months to years \citep{2004VandenBerk, MacLeod2010}, and such variations could naturally lead to small changes in continuum brightness and line profiles between non-simultaneous observations. Meanwhile, differences in instrumental response, spectral resolution, and observational setup between BOSS and DESI may also partly account for the observed differences.
    We also examine the flux ratio between the two members (Figure~\ref{fluxratio}), which remains nearly constant within the uncertainties, providing additional evidence for a lensing interpretation.
    
    Although the separation of this system is $8.98\arcsec$, which is larger than the typical separations ($\lesssim4\arcsec$) known for galaxy-scale lensed quasars, a single massive galaxy can produce image separations up to $\sim10.1\arcsec$ \citep{Stern2021ApJ}. 
    Therefore, J0023+0417 remains a potential lensed system. As part of our ongoing work, we are carrying out spectroscopic observations to determine the redshift of the central red source and confirm its nature. After obtaining the spectra, we will perform lens mass modeling to better constrain the mass distribution and overall properties of the lens system.
    
    \section{Summary} \label{sec:sum}
    In this work, we retrieved spectra for MGQPC candidates from SPARCL and applied a conservative association criterion of $\lvert \Delta v \rvert \leqslant 2000\ \mathrm{km\,s^{-1}}$. We confirmed 17 dual quasars and 143 projected quasars. The dual quasar sample spans z=0.573--2.758 (nine systems at $z>1.5$). Although spectral differences could, in principle, arise from differential extinction, chromatic microlensing, or time-delay variability, the lack of an obvious deflector/environment in the available image and the observed spectral non-homologies support a dual quasar scenario. We calculate the dual quasar fraction which remains at the level of $\sim 10^{-4}$ over most of the redshift range 0.5 < z < 3.5, with only a slight enhancement around z $\sim$ 1 to 2.5.  The projected quasars sample spans z=0.324–4.030 with $r_p$=17.467--99.946 kpc and includes 16 particularly close systems ($r_p\lesssim30$ kpc) that offer promising foreground–background sightlines for future absorption-line studies of gas around foreground quasar hosts, while a subset with high-$z$ backgrounds naturally provides Ly$\alpha$ forest coverage for potential IGM tests. We found extra five pairs, for which \texttt{member\_A} has only photometric redshifts, and are therefore classified as dual quasar candidates.
    Moreover, our search had discovered one strong lensing candidate, J0023+0417. We will be pursued high-resolution spectra and detailed lens mass modeling to confirm or rule out its lensing nature.

\onecolumn
    \begin{longtable}{lrrrrrrr}
    \caption{The basic features of 143 projected quasars. $r_p$ is the impact parameter. (This entire table is available in machine-readable form.)
    }\\
    \label{PQ_parameters}\\
    \hline\hline
    Name$^{\,(1)}$ & R.A.$^{\,(2)}$ & Dec$^{\,(3)}$ & G$^{\,(4)}$ & Redshift$^{\,(5)}$ & SepAB$^{\,(6)}$ & $r_p$$^{\,(7)}$ & $\lvert \Delta v_r \rvert$$^{\,(8)}$\\
\hline
\endfirsthead
\caption{continued.}\\
\hline
Name$^{\,(1)}$ & R.A.$^{\,(2)}$ & Dec$^{\,(3)}$ & G$^{\,(4)}$ & Redshift$^{\,(5)}$ & SepAB$^{\,(6)}$ & $r_p$$^{\,(7)}$ & $\lvert \Delta v_r \rvert$$^{\,(8)}$ \\
\hline
\endhead
\hline
\endfoot

J0020+1932A & 5.117593 & 19.536961 & 19.11 & 1.20675 & \multirow{2}{*}{9.477} & \multirow{2}{*}{76.794} & \multirow{2}{*}{21\,241.924} \\
J0020+1932B & 5.119607 & 19.538783 & 20.84 & 1.05574 &  &  &  \\
\hline

J0033+1604A & 8.368102 & 16.078592 & 20.52 & 2.44169 & \multirow{2}{*}{8.185} & \multirow{2}{*}{65.953} & \multirow{2}{*}{154\,140.280} \\
J0033+1604B & 8.366692 & 16.080414 & 20.10 & 1.03401 &  &  &  \\
\hline

J0036+0747A & 9.215690 & 7.786463 & 20.52 & 1.29287 & \multirow{2}{*}{8.049} & \multirow{2}{*}{65.913} & \multirow{2}{*}{24\,675.527} \\
J0036+0747B & 9.217925 & 7.786155 & 19.71 & 1.11161 &  &  &  \\
\hline

J0049+0127A & 12.347273 & 1.452848 & 20.70 & 1.06040 & \multirow{2}{*}{10.124} & \multirow{2}{*}{82.131} & \multirow{2}{*}{65\,167.703} \\
J0049+0127B & 12.350059 & 1.453240 & 20.70 & 1.56289 &  &  &  \\
\hline

J0106$-$0724A & 16.725155 & $-$7.410522 & 19.89 & 3.13129 & \multirow{2}{*}{11.123} & \multirow{2}{*}{84.571} & \multirow{2}{*}{180\,165.700} \\
J0106$-$0724B & 16.728269 & $-$7.410428 & 19.75 & 1.22217 &  &  &  \\
\hline

J0110+1030A & 17.624062 & 10.503477 & 19.62 & 2.06080 & \multirow{2}{*}{10.223} & \multirow{2}{*}{71.329} & \multirow{2}{*}{180\,279.780} \\
J0110+1030B & 17.626175 & 10.505454 & 17.21 & 0.64568 &  &  &  \\
\hline

J0118+2005A & 19.679380 & 20.089000 & 20.78 & 2.62275 & \multirow{2}{*}{10.275} & \multirow{2}{*}{82.047} & \multirow{2}{*}{88\,156.305} \\
J0118+2005B & 19.680166 & 20.091760 & 20.56 & 1.69400 &  &  &  \\
\hline

J0126+1539A & 21.595740 & 15.653657 & 20.64 & 1.06601 & \multirow{2}{*}{3.801} & \multirow{2}{*}{30.963} & \multirow{2}{*}{5\,338.558} \\
J0126+1539B & 21.595340 & 15.652671 & 20.18 & 1.10313 &  &  &  \\
\hline

    \hline
        \end{longtable}

\twocolumn

\section*{Data Availability}
The catalogs of the 17 DQs, 143 PQs, and an LQ candidate are available at the CDS via \url{https://cdsarc.cds.unistra.fr/viz-bin/cat/J/A+A/XXXXX}.

\begin{acknowledgements}
This work has been supported by the National Key R\&D Program of China (2025YFA1614101, 2021YFA0718500), and by the Chinese National Natural Science Foundation grants 12333001. 

Guoshoujing Telescope (the Large Sky Area Multi-Object Fiber Spectroscopic Telescope, LAMOST) is a National Major Scientific Project built by the Chinese Academy of Sciences. Funding for the project has been provided by the National Development and Reform Commission. LAMOST is operated and managed by the National Astronomical Observatories, Chinese Academy of Sciences.

This research uses services or data provided by the SPectra Analysis and Retrievable Catalog Lab (SPARCL) and the Astro Data Lab, which are both part of the Community Science and Data Center (CSDC) program at NSF's National Optical-Infrared Astronomy Research Laboratory. 
NOIRLab is operated by the Association of Universities for Research in Astronomy (AURA), Inc. under a cooperative agreement with the National Science Foundation.

The DESI Legacy Imaging Surveys consist of three individual and complementary projects: the Dark Energy Camera Legacy Survey (DECaLS), the Beijing-Arizona Sky Survey (BASS), and the Mayall z-band Legacy Survey (MzLS). Pipeline processing and analyses of the data were supported by NOIRLab and the Lawrence Berkeley National Laboratory (LBNL). Legacy Surveys was supported by: the Director, Office of Science, Office of High Energy Physics of the U.S. Department of Energy; the National Energy Research Scientific Computing Center, a DOE Office of Science User Facility; the U.S. National Science Foundation, Division of Astronomical Sciences; the National Astronomical Observatories of China, the Chinese Academy of Sciences and the Chinese National Natural Science Foundation. LBNL is managed by the Regents of the University of California under contract to the U.S. Department of Energy.
\end{acknowledgements}

\bibliographystyle{aa}
\bibliography{ref}

\end{document}